\documentclass[aps,prb,preprint,eqsecnum,showpacs,showkeys,byrevtex,tightenlines]{revtex4}
\begin{document}
\title{Spectral Zeta Functions for
       a Cylinder and a Circle}
\author{V.~V.~Nesterenko\thanks{E-mail: nestr@thsun1.jinr.ru},
I.~G.~Pirozhenko\thanks{E-mail: pirozhen@thsun1.jinr.ru}}
\address{Bogoliubov Laboratory of Theoretical Physics \\
Joint Institute for Nuclear Research, Dubna, 141980, Russia}
%\author{M.~Bordag\thanks{E-mail: Michael.Bordag@itp.uni-leipzig.de}}
%\address{Universit\"at Leipzig, Institute f\"ur Theoretical Physik \\
%Augustusplatz 10, 04109 Leipzig, Germany}
\date{\today}
\begin{abstract}
Spectral zeta functions $\zeta(s)$ for the massless scalar fields obeying the
Dirichlet and Neumann boundary conditions on a surface of an infinite
cylinder are constructed.  These functions are defined
explicitly in a finite domain of the complex plane $s$ containing the
closed interval of real axis
$-1\le$ Re\,$s \le 0$. Proceeding from this
the spectral zeta functions for the boundary conditions given on a circle
(boundary value problem on a plane) are obtained without any additional
calculations. The Casimir energy for the relevant
field configurations is deduced.
\end{abstract}
\thispagestyle{empty}
\pacs{12.20.Ds, 03.70.+k, 12.40.-y}
\maketitle
\section{Introduction}

The zeta function technique  is widely used in quantum field
theory, specifically, for calculation of the vacuum energy
(the Casimir energy) of  quantized fields in  compactified
configuration space.~\cite{Birel-Davis,Odintsov,Elizalde}
For the boundaries with spherical geometry this method has
been essentially worked out in papers.~\cite{new}
The Casimir energy is determined by the value of the corresponding
zeta function $\zeta(s)$ at a certain point (usually at $s=-1$).
Therefore, when it regards the vacuum energy calculation,
the zeta function is investigated only at a separate point or at least in
its infinitesimal neighborhood.

However, in some problems it proves to be useful to construct
explicitly the spectral zeta function for a finite range of its
argument $s$. For example, proceeding from the zeta function for an infinite
cylinder $\zeta_{\text{cyl}}(s)$  defined in the domain  $-1\le$ Re~$s\le 0$
one can easily express both the Casimir energy of a cylinder via
$\zeta_{\text{cyl}}(-1)$ and the Casimir energy of a circle in terms of
$\zeta_{\text{cyl}}(0)$. Such an approach has obvious advantage
not only for shortening the calculations
(a unique  zeta function is applicable to two problems)
but also when treating the divergences.
For instance, the zeta function technique applied to the
boundary conditions given  on the surface of an infinite cylinder at once
gives the finite value of the vacuum energy~\cite{LNB,MNN} (the
renormalization is carried out simultaneously with the regularization).
As for the boundary conditions  defined on a circle, the
zeta function regularization is unable to remove all the divergences.
The answers obtained by different authors for the finite part of
the corresponding Casimir energy  do not coincide.~\cite{L+Romeo}
In view of this it would be interesting to express the spectral zeta function
$\zeta_{\text{cir}}(s)$
in terms of $\zeta_{\text{cyl}}(s)$ which is  already "normalized" (being free of the
divergences)  at the point $s=-1$.  It is this that  we are going to do
in the present note.

The layout of the paper is as  follows.
In Sec.~II we derive  the
relation between the zeta functions for the boundary conditions
given on the surface of an infinite cylinder, $\zeta_{\text{cyl}}(s)$,
and on the circle, $\zeta_{\text{cir}}(s)$. In Sec.~III the spectral zeta functions
for the massless  scalar fields obeying the Dirichlet and Neumann
boundary conditions on the lateral area of an infinite cylinder are
constructed explicitly. As in Ref.~\onlinecite{LNB} the central part is played
here by the
uniform asymptotic expansion of the Bessel functions. Explicit
formulas defining these zeta functions in a finite region $\Omega$
of the complex plane $s$ containing the closed interval  of real axis
$-1\le$ Re~$s\le 0$ are derived. In Sec.~IV using these
formulas we obtain the
spectral zeta functions for the relevant plane problem
with the Dirichlet or Neumann boundary conditions given on a circle.
The Casimir energies for the  field configurations
in hand are calculated. In Conclusion (Sec.~V)
the obtained results are shortly discussed and compared with those
of other authors.

\section{Relation between the spectral zeta functions for a cylinder and
 a circle}

Let us remind briefly how to find out the relation between the zeta
function   for a three dimensional problem with boundary conditions given
 on the lateral area of an infinite cylinder of radius $a$ and the
zeta function for a two dimensional  problem with boundary conditions
given on a circle of the same radius.  For simplicity in both cases
the massless scalar field obeying the Dirichlet or Neumann boundary
conditions is considered.

In the case of an infinite cylinder the eigenfunctions
are proportional to $\exp (-i\omega t+i k_z z +i n \theta)$
where  $\{r, \theta, z\}$ are cylindrical coordinates.
The eigenfrequencies $\omega$ inside the cylinder for the Dirichlet
and Neumann boundary conditions  are given  respectively by the
equations
\begin{equation}
\left. J_l(\lambda r)\right|_{r=a}=0\,{,} \quad
\left. J'_l(\lambda r)\right|_{r=a}=0\,{.}
\label{e_2_1}
\end{equation}
For the outside region we have
\begin{equation}
\left. H^{(1)}_l(\lambda r)\right|_{r=a}=0\,{,} \quad
\left.{H^{(1)}_l}'(\lambda r)\right|_{r=a}=0\,{.}
\label{e_2_2}
\end{equation}
Here the notation  $\lambda^2=\omega^2-k_z^2$ is introduced.

The eigenfrequency equations for the scalar field with boundary
conditions defined on a circle are obtained by putting $k_z^2=0$
in (\ref{e_2_1}) and (\ref{e_2_2}).

In order to construct the spectral zeta function for an infinite
 cylinder or a
circle one can  employ the standard definition
\begin{equation}
\zeta (s)=\sum_{\{p\}}(\omega_p^{-s}-\bar {\omega}_p^{-s})\,{.}
\label{e_2_3}
\end{equation}
Here $\omega_p$  are the eigenfrequencies of the scalar field under certain
boundary conditions, $\tilde{\omega}_p$ are the same frequencies
when the boundaries are removed.
The summation (or integration) should be done over all the quantum
numbers $\{p\}$ specifying the spectrum.
To make the sum convergent the parameter $s$ should belong to
a region of the complex plane $s$ where Re~$s$ is large enough. However
for the  massless fields considered here there also exists
a restriction for the maximal values of  Re~$s$ in order to ensure
the convergent integration at the origin  (see below).

For a cylinder  and a circle the general formula (\ref{e_2_3})
looks as follows
\begin{eqnarray}
\zeta_{\text{cyl}}(s)&=&\int_{-\infty}^{\infty}\frac{d k_z}{2 \pi}
\sum \limits_{l,n}^{}\left[(\lambda^2_{ln}(a)+k_z^2)^{-s/2}-
(\lambda^2_{ln}(\infty)+k_z^2)^{-s/2}\right],
\label{e_2_4}\\
\zeta_{\text{cir}}(s)&=&\sum \limits_{l,n}^{}\left[\lambda^s_{ln}(a)-
\lambda^s_{ln}(\infty)\right],
\label{e_2_5}
\end{eqnarray}
with $\lambda_{ln}$ being defined by (\ref{e_2_1}) and (\ref{e_2_2}) for
the both
zeta functions.
Integration over $k_z$ in (\ref{e_2_4}) can be accomplished
by making use of the formula
$$
\int_{-\infty}^{\infty} \frac{d k_z}{2 \pi} (k_z^2+b^2)^{-s/2}=
\frac{b^{1-s}}{2 \pi}\,B\left(\frac{1}{2},\frac{s-1}{2}\right),\quad
\mbox{Re } s>1\,{,}
$$
where $B(x, y)$  is the Euler beta function
$$
B(x,y)=\Gamma(x) \Gamma(y)/\Gamma(x+y).
$$
Comparing the result of the integration with (\ref{e_2_5}) one arrives
at the  relation between $\zeta_{\text{cyl}}(s)$ and
$\zeta_{\text{cir}}(s)$~\cite{GR1}
\begin{equation}
\zeta_{\text{cyl}}(s)=\frac{1}{2 \pi} B\left(\frac{1}{2}, \frac{s-1}{2}\right)
\zeta_{\text{cir}}(s-1)\,{.}
\label{e_2_6}
\end{equation}
When calculating the Casimir energy by making use of the zeta function
technique  usually one puts~\cite{Nucl-Phys}
\begin{equation}
E_{\text{C}}=\frac{1}{2}\zeta(s=-1).
\label{e_2_7}
\end{equation}
To use  this formula, for example,  in the case of a circle
one should find the analytic continuation  of
$\zeta_{\text{cir}}(s)$ into the point $s=-1$. On the other hand, in accordance
with~(\ref{e_2_6}) the zeta function  $\zeta_{\text{cir}}(s)$ at the point
$s=-1$ can be expressed through $\zeta_{\text{cyl}}(s=0)$.
Thus the analytic continuation of the zeta function
$\zeta_{\text{cyl}}(s)$   into the region  $-1\le$ Re~$s\le 0$  provides
the opportunity to calculate the Casimir energy both for an infinite cylinder
and for a circle.

\section{Spectral zeta functions for a cylinder with the Dirichlet
and Neumann boundary conditions}

In Ref.~\onlinecite{LNB} a consistent procedure has been developed for
constructing the spectral zeta functions for the boundary conditions
given on a sphere and on the
lateral area of an infinite cylinder. Here we follow
the same approach and start with consideration of the spectral zeta
function  $\zeta_{\text{cyl}}(s)$ for the massless scalar field obeying the
Dirichlet boundary conditions on an infinite cylinder.

Taking into account the contributions of the field oscillations
inside (Eq.~(\ref{e_2_1})) and
outside (Eq.~(\ref{e_2_2})) the cylinder and representing the sum over $l$
in~(\ref{e_2_4}) in terms of contour integral one obtains
\begin{equation}
\zeta_{\text{cyl}}^{\text{D}}(s)=\frac{1}{2\pi}\int_{-\infty}^{\infty}
\frac{d k_z}{2 \pi i}\sum\limits_{n=-\infty}^{\infty}
\oint_{C}^{}(\lambda^2+k_z^2)^{-s/2}d_{\lambda}
\ln\frac{J_n(\lambda a) H_n^{(1)}(\lambda a)}{J_n(\infty)
H_n^{(1)}(\infty)}.
\label{e_3_1}
\end{equation}
The contour $C$ consists of the imaginary axis $(-i\infty,i\infty)$
and a semi-circle  of an infinite radius in the right half plane
of a complex variable $\lambda$. Keeping in mind the behavior of the
integrand  at all the segments  of the contour $C$
and integrating over $k_z$ Eq.~(\ref{e_3_1}) becomes~\cite{LNB}
\begin{equation}
\zeta_{\text{cyl}}^{\text{D}}(s)=\frac{a^{s-1}}{2\sqrt{\pi}
\Gamma\left(\frac{\displaystyle s}
{\displaystyle 2}\right)\Gamma\left(\frac{\displaystyle 3-s}
{\displaystyle 2}\right)}
\sum\limits_{n=-\infty}^{\infty}\int_{0}^{\infty}
dy\, y^{1-s}\frac{d}{dy}\ln\left[2y I_n(y) K_n(y)\right].
\label{e_3_2}
\end{equation}
Then, in order to accomplish the analytic  continuation  of~(\ref{e_3_2})
into the region $-1\le$ Re~$s<0$  we express $\zeta_{\text{cyl}}(s)$ in terms of
the Riemann zeta function with the well-known analytic continuation.
After changing the integration variable $y\to ny$ in (\ref{e_3_2})
we employ  the uniform asymptotic expansion (UAE) of the Bessel
functions~\cite{AS} up to the order $n^{-4}$
\begin{eqnarray}
\ln\left(2 y n I_n(ny)K_n(ny)\right)&=&
\ln(y t)+\frac{t^2(1-6t^2+5t^4)}{8 n^2}\nonumber\\
&&+\frac{t^4(13-284 t^2+ 1062 t^4-1356 t^6+ 565 t^8)}{64 n^4}+O(n^{-6})\,{,}
\label{e_3_3}
\end{eqnarray}
where $t=1/\sqrt{1+y^2}$.
Substituting  (\ref{e_3_3}) in all the terms of the series (\ref{e_3_2}),
where $n\ne0$, we obtain
\begin{eqnarray}
\zeta^{\text{D}}_{\text{cyl}}(s)&=&C(s)\left(Z_0(s)+Z_1(s)+Z_2(s)+Z_3(s)\right),
\label{e_3_4}\\
Z_0(s)&=&\int_0^{\infty} dy\, y^{1-s}\frac{d}{dy}
\left\{\ln(2y I_0(y) K_0(y))-\frac{t^2}{8}(1-6 t^2+5 t^4)\right\},
\label{e_3_5}\\
Z_1(s)&=&\sum\limits_{n=1}^{\infty}n^{1-s}\int_{0}^{\infty}
dy\, y^{1-s}\frac{d}{dy}\ln \left(\frac{y^2}{1+y^2}\right),
\label{e_3_6}\\
Z_2(s)&=&\frac{1}{4}\left(\sum\limits_{n=1}^{\infty}n^{-1-s}+
\frac{1}{2}\right)\int_{0}^{\infty}
dy\, y^{1-s}\frac{d}{dy}\left[t^2(1-6\, t^2+5\, t^4)\right],
\label{e_3_7}\\
Z_3(s)&=&\frac{1}{32}\sum\limits_{n=1}^{\infty}n^{-3-s}
\int_{0}^{\infty}
dy\, y^{1-s}\frac{d}{dy}\left[t^4(13-284\, t^2+1062\, t^4-1356\, t^6
+565\, t^8)\right],
\label{e_3_8}
\end{eqnarray}
where
\begin{equation}
C(s)=\frac{a^{s-1}}{2\sqrt{\pi}\Gamma\left(\frac{\displaystyle s}
{\displaystyle 2}\right)
\Gamma\left(\frac{\displaystyle 3-s}{\displaystyle 2}\right)}\,{.}
\label{e_3_9}
\end{equation}
Here the notation $Z_0(s)$ is introduced for the difference between
the term with $n=0$ in (\ref{e_3_2}) and the integral
\begin{equation}
A(s)=\frac{1}{8}\int_0^{\infty}dy\, y^{1-s}\frac{d}{dy}
\left[t^2(1-6\, t^2+5\, t^4)\right], \quad -1<\mbox{Re }s<3.
\label{e_3_10}
\end{equation}
The function $Z_1(s)$ corresponds to the first term in the UAE (\ref{e_3_3}),
$Z_2(s)$  involves the contribution  of the $1/n^2$-order term
of the latter together with the integral $A(s)$, $Z_3(s)$ is generated
by the terms of order $1/n^4$  in the expansion (\ref{e_3_3}).

Taking into account  the asymptotics
\begin{eqnarray}
2y I_0(y) K_0(y)&=&-2y\ln y+(2\ln 2-2 \gamma)y +O(y^3), \quad y\to 0\,{,}
\nonumber\\
2y I_0(y) K_0(y)&=&1+\frac{1}{8\, y^2}+\frac{27}{128\, y^4}+O(y^{-6}),\quad
y\to\infty\,{,}
\label{e_3_11}
\end{eqnarray}
where $\gamma$ is the Euler constant, one can ascertain  the domain
of variation of the complex variable  $s$  so that the integrals
in (\ref{e_3_5})--(\ref{e_3_8}) exist.
The ultraviolet behavior $(y\to\infty)$ of the integrands
in (\ref{e_3_5})--(\ref{e_3_8})  determines
the lower bound for Re~$s$
and the infrared one  $(y\to0)$ is responsible for the upper bound.
Formula  (\ref{e_3_5}) defines
$Z_0(s)$ as an analytic function of $s$  if $-3<$ Re~$s<1$.
When this condition holds one can perform in (\ref{e_3_5}) the
integration by parts
\begin{equation}
Z_0(s)=-(1-s)\int_0^{\infty}dy\, y^{-s}\left[\ln(2y I_0(y)K_0(y))-
\frac{t^2}{8}(1-6t^2+5t^4)\right].
\label{e_3_12}
\end{equation}

The integral defining $Z_1(s)$ in (\ref{e_3_6}) exists if  $-1<$ Re~$s<1$.
The sum over $n$ in  this formula is finite for Re~$s > 2$. As these two
regions do not overlap, the introduction of the parameter $s$
in the original formula
(\ref{e_2_3}) does not regularize completely the divergences in $Z_1(s)$
on this stage of our consideration.
An additional infrared regularization should be used here,
for example, by introducing the photon "mass" $\mu $. As a result the
integration in Eq.~(\ref{e_3_6}) will be restricted from below by $\mu $, and
the constraint
Re~$s<1$ will be removed. The function $Z_1(s)$, regularized in this way,
can be used for required analytic continuation (see below).

The integral in Eq.~(\ref{e_3_7}) defining  $Z_2(s)$
exists when $-1<$ Re~$s<3$. The sum over $l$ in (\ref{e_3_7})
is finite when Re~$s>0$.
The integral in (\ref{e_3_8}) converges  if  $-3<$ Re~$s<3$. The sum over
$n$ in this formula is finite  for Re~$s>-2$. Thus the regions, where
the integrals and the sums exist, overlap and  these formulas can be used
for constructing the analytic continuation needed.

As it was stressed in the preceding section, we are interested in
the evaluation of  $\zeta^{\text{D}}_{\text{cyl}}(s)$  in the region  $\Omega $
of the complex plane $s$
containing  the closed interval of the real axis  $-1\le$ Re~$s\le 0$.
For the  zeta function  $\zeta^{\text{D}}_{\text{cyl}}(s)$ defined by the sum
(\ref{e_3_4}) we shall construct the analytic continuation  into
this region in the following way.

It has been already noticed that the functions $Z_0(s)$ in (\ref{e_3_5})
and $Z_3(s)$ in (\ref{e_3_8}) are analytic in the  region under consideration.
In order to obtain  the required  analytic continuation  of the
function $Z_2(s)$  it as sufficient to express the sum over  $n$
in~(\ref{e_3_7}) in terms of the Riemann zeta function
\begin{equation}
\sum\limits_{n=1}^{\infty}\frac{1}{n^z}=\zeta(z),
\label{e_3_13}
\end{equation}
and the integral  in (\ref{e_3_7}) in terms of the Euler gamma function
using the equality
\begin{equation}
\int_0^{\infty}dy \,y^{1-s}\frac{d}{dy}t^{2(\rho-1)}=
(1-\rho)\,\frac{\Gamma\left(\frac{\displaystyle 3-s}{\displaystyle
 2}\right)
\Gamma\left(\rho-\frac{\displaystyle  3-s}{\displaystyle  2}
\right)}{\Gamma(\rho)},
\quad 3-2\mbox{ Re }\rho<\mbox{ Re }s<3.
\label{e_3_14}
\end{equation}
It gives
\begin{equation}
Z_2(s)=\frac{1}{4}\left[\zeta(s+1)+\frac{1}{2}\right]
\Gamma\left(\frac{3-s}{2}\right)\Gamma\left(\frac{1+s}{2}\right)
\left[-1+3(1+s)-\frac{5}{8}(3+s)(1+s)\right].
\label{e_3_15}
\end{equation}
Making use of Eq.\ (\ref{e_3_14})  for integrating in  (\ref{e_3_8})
we get
\begin{eqnarray}
Z_3(s)&=&\frac{1}{32}\zeta(s+3)\Gamma\left(\frac{3-s}{2}\right)
\left[-13\, \Gamma\left(\frac{3+s}{2}\right)+
142\, \Gamma\left(\frac{5+s}{2}\right)-177\,
\Gamma\left(\frac{7+s}{2}\right)\right.\nonumber\\
&&\left.+\frac{113}{2}\,
\Gamma\left(\frac{9+s}{2}\right)-\frac{113}{24}\,
\Gamma\left(\frac{11+s}{2}\right)\right].
\label{e_3_16}
\end{eqnarray}

Keeping in mind that we have introduced the photon mass
$\mu $ into Eq.~(\ref{e_3_6}) we can substitute here the sum in terms of
the Riemann zeta function according to Eq.~(\ref{e_3_13}). After that the
photon mass can be put to zero.
In order to obtain the required analytic
continuation of the integral in~(\ref{e_3_6})
we first expand the logarithm
\begin{equation}
\ln\frac{y^2}{1+y^2}=\ln\left(1-\frac{1}{1+y^2}\right)
=-\sum\limits_{m=1}^{\infty}\frac{1}{m(1+y^2)^m}.
\label{e_3_17}
\end{equation}
After that  one can  carry out the integration in (\ref{e_3_6})
with the result
\begin{equation}
Z_1(s)=\frac{1-s}{2} \zeta(s-1)\Gamma\left(-\frac{1-s}{2}\right)
\sum\limits_{m=1}^{\infty}\frac{1}{m}
\frac{\Gamma\left(m-\frac{\displaystyle  1-s}{\displaystyle
 2}\right)}{\Gamma(m)}.
\label{e_3_18}
\end{equation}
In order for the domain of the convergence of the series in
(\ref{e_3_18}) to be determined it
is convenient to use the formula 8.328.2 from Ref.~\onlinecite{GR}
\begin{equation}
\left.\frac{\Gamma(m+z)}{\Gamma(m)}\right|_{m\to\infty}\to
\frac{1}{m^{(1-z)/2}}\,{.}
\label{e_3_19}
\end{equation}
From Eq.\ (\ref{e_3_19}) it follows that the series (\ref{e_3_18}) converges
when Re~$s<1$. But it is not dangerous now because we have replaced the sum
in Eq.~(\ref{e_3_6}) in terms of the Riemann zeta function $\zeta (s-1)$
which is defined everywhere except for the point $s = 2$.

Finally
Eqs.\  (\ref{e_3_4}), (\ref{e_3_9}), (\ref{e_3_12}),
(\ref{e_3_15}), (\ref{e_3_16}), and (\ref{e_3_18}) define
the spectral zeta function  $\zeta^{\text{D}}_{\text{cyl}}(s)$ as an analytic
function of the complex variable $s$ in the region~$\Omega $.
Here it is worth noting  that the analytic  continuation
does not ensure that in the region $\Omega$, we are interested in,
the spectral zeta function $\zeta_{\text{cyl}}(s)$ is free of singularities.
Really Eqs.\ (\ref{e_3_13}) and (\ref{e_3_14}) used for analytic continuation
contain the Riemann zeta function  $\zeta(s)$ and the Euler gamma
function $\Gamma(s)$ having
singularities (poles)  at certain isolated points. Therefore,
$\zeta_{\text{cyl}}^{\text{D}}(s)$ considered in $\Omega$
may also possess the singularities of the same type.~\cite{endnote}
As the  analytic continuation is unique the removal of these divergences
is obviously impossible. This manifests the inability of the present
approach to remove  the divergences in all the cases. If the
considered quantity is expressed via the value of the spectral zeta function
at its singularity
point then the zeta function technique  does not give a finite
answer.~\cite{Nucl-Phys}

In order to obtain the Casimir energy of the massless scalar field
obeying the Dirichlet boundary conditions on an infinite cylinder
of radius $a$  let us calculate, according to (\ref{e_2_7}),
the spectral zeta function $\zeta_{\text{cyl}}^{\text{D}}(s)$ at the point  $s=-1$.
Numerical integration in (\ref{e_3_12}) yields
\begin{equation}
Z_0(-1)=-0.021926+\frac{3}{4}-\frac{5}{16}=0.415574.
\label{e_3_20}
\end{equation}

Now we turn to Eq.\ (\ref{e_3_18}) with  $s$ tending  to~$-1$
\begin{equation}
Z_1(-1)=\lim_{s\to-1}\zeta(s-1)\left[\Gamma\left(\frac{1+s}{2}\right)
+\sum\limits_{m=2}^{\infty}\frac{1}{m(m-1)}\right].
\label{e_3_21}
\end{equation}
Keeping   in mind the relations
\begin{eqnarray}
\Gamma(x)&=&\frac{1}{x}-\gamma+O(x),\nonumber\\
\sum\limits_{m=2}^{\infty}\frac{1}{m(m-1)}&=&1,\quad \zeta(-2)=0,
\label{e_3_22}
\end{eqnarray}
where $\gamma $ is the Euler constant, $\gamma =0.577215\ldots$,
one derives
\begin{equation}
Z_1(-1)=2 \lim_{s\to-1}^{}\frac{\zeta(s-1)}{1+s}+
\zeta(-2)(1-\gamma)=2\, \zeta'(-2)=-0.060897\,{.}
\label{e_3_23}
\end{equation}

Now we evaluate  the value of $Z_2(-1)$ using Eq.\  (\ref{e_3_15})
and taking into account that $\Gamma((1+s)/2)$ has a pole at the point
$s=-1$ (see Ref.~\onlinecite{LNB})
\begin{eqnarray}
Z_2(-1)&=&-\frac{1}{4}\lim_{s\to-1}^{}
\left[\zeta(s+1)+\frac{1}{2}\right]\Gamma\left(\frac{1+s}{2}\right)
\nonumber\\
&=&-\frac{1}{4}\lim_{s\to-1}^{}\left[\zeta(0)+\zeta'(0)(s+1)
+O\left((s+1)^2\right)+\frac{1}{2}\right]\cdot
\left[\frac{2}{1+s}-\gamma+O\left(\frac{1+s}{2}\right)\right]
\nonumber\\
&=&-\frac{1}{4}2\,\zeta'(0)=-\frac{1}{4}\ln(2\pi)\,{.}
\label{e_3_24}
\end{eqnarray}
Here we have used  the values of the Riemann zeta function and
its derivative at the origin
\begin{equation}
\zeta(0)=-\frac{1}{2},\quad \zeta'(0)=-\frac{1}{2}\ln(2\pi).
\label{e_3_25}
\end{equation}

The formula (\ref{e_3_16}) gives the following value for $Z_3(-1)$
\begin{equation}
Z_3(-1)=\frac{1}{32}\zeta(2)=\frac{\pi^2}{192}=0.051404\,{.}
\label{e_3_26}
\end{equation}
One can make this result more precise.
The point is that, from the very beginning, a complete
expression
\begin{equation}
\bar{Z}_3(s)=2\sum_{n=1}^{\infty}n^{1-s}
\int_0^{\infty} dy\, y^{1-s}\frac{d}{dy}
\left[\ln(2yn \,I_n(yn) K_n(ny))-
\ln\frac{y}{\sqrt{1+y^2}}-\frac{t^2(1-6 t^2+5 t^4)}{8 n^2}\right]
\label{e_3_27}
\end{equation}
can be considered instead of the function $Z_3(s)$ defined by (\ref{e_3_8}).
The formula (\ref{e_3_27}) reduces to  (\ref{e_3_8}) after substituting the logarithm
by its uniform asymptotic expansion~(\ref{e_3_3}).
It is easy to show that $\bar{Z}_3(s)$ is an  analytic function of $s$
when $-3<$ Re~$s<1$. It means that $\bar{Z}_3(s)$, as well as $Z_3(s)$,
does not need analytic continuation. In practice Eq.\  (\ref{e_3_27})
is used for several first values of $n$,  $n\le n_0$,  and for  $n>n_0$
one applies (\ref{e_3_8}). The reason is that the uniform asymptotic
expansion does not
provide sufficient accuracy when $n<n_0=6\div 10$. Evaluating  $Z_3(-1)$
according to this algorithm with $n_0=6$ we obtain an improved
value (compare with (\ref{e_3_26}))
\begin{equation}
\bar{Z}_3(-1)=0.045611\,{.}
\label{e_3_28}
\end{equation}
For the remaining coefficient $C(s=-1)$ in (\ref{e_3_4}) one finds
\begin{equation}
C(-1)=-\frac{1}{4\pi a^2}.
\label{e_3_29}
\end{equation}

Finally, summing up Eqs.~(\ref{e_3_20}), (\ref{e_3_23}), (\ref{e_3_24}),
(\ref{e_3_28}), and (\ref{e_3_29})  we get  for $\zeta_{\text{cyl}}^{\text{D}}(-1)$
\begin{equation}
\zeta_{\text{cyl}}^{\text{D}}(-1)=-\frac{1}{4\pi a}\left[0.415574-0.060897-
\frac{1}{4}\ln(2\pi)+0.045611 \right]=\frac{0.001213}{a^2}\,{.}
\label{e_3_30}
\end{equation}
It gives the following value for the Casimir energy of massless scalar
field obeying the Dirichlet boundary conditions on the lateral area of
an infinite cylinder of radius~$a$
\begin{equation}
E_{\text{cyl}}^{\text{D}}=\frac{1}{2}
\zeta_{\text{cyl}}^{\text{D}}(-1)=\frac{0.000606}{a^2}.
\label{e_3_31}
\end{equation}

It is not necessary to calculate  the spectral zeta function for the
Neumann boundary conditions  $\zeta_{\text{cyl}}^{\text{N}}(s)$.
 The point is that in
Ref.~\onlinecite{LNB} the spectral zeta function for the electromagnetic
field with boundary
conditions  defined on an infinitely thin perfectly
conducting cylindrical shell was constructed. This zeta function
is the sum of  two spectral  zeta functions for scalar fields
obeying the
Dirichlet  and Neumann  boundary conditions on a cylinder.
Therefore
\begin{equation}
\zeta^{\text{N}}_{\text{cyl}}(s)=\zeta_{\text{cyl}}^{\text{shell}}(s)
-\zeta_{\text{cyl}}^{\text{D}}(s).
\label{e_3_32}
\end{equation}
We shall not quote here the expression for $\zeta_{\text{cyl}}^{\text{shell}}(s)$
found in Ref.~\onlinecite{LNB} (see the next section). Taking into account
Eq.~(\ref{e_3_32}) at $s=-1$
we obtain the Casimir energy  of massless scalar field with Neumann
boundary conditions on an infinite cylinder
\begin{equation}
E^{\text{N}}=E_{\text{cyl}}^{\text{EM}}-E_{\text{cyl}}^{\text{D}}=-\frac{0.01356}{a^2}-\frac{0.00061}{a^2}
=-\frac{0.01417}{a^2}\,{.}
\label{e_3_33}
\end{equation}
Here we have borrowed the value of the Casimir energy  $E_{\text{cyl}}^{\text{shell}}$
from Ref.~\onlinecite{MNN} where its consistent derivation is presented.
Recently~\cite{GR1}   the values of the zeta  functions
$\zeta_{\text{cyl}}^{\text{D}}(-1)$
and $\zeta_{\text{cyl}}^{\text{N}}(-1)$ were evaluated with higher accuracy
(see the Conclusion).

\section{Spectral zeta functions for a circle}
Having defined the zeta functions for an infinite
cylinder $\zeta_{\text{cyl}}^{\text{D}}(s)$  and
$\zeta_{\text{cyl}}^{\text{N}}(s)$ in
the region $\Omega $ of the complex plane~$s$, containing
the segment of the real axis $-1\le$ Re~$s\le 0$, we can at once obtain
the zeta functions for a circle, $\zeta_{\text{cir}}^{\text{D}}(s)$
and $\zeta_{\text{cir}}^{\text{N}}(s)$,  making use of Eq.\ (\ref{e_2_6})
\begin{equation}
\zeta_{\text{cir}}^{\text{D,N}}(s)=2\sqrt{\pi}
\frac{\Gamma\left(\frac{\displaystyle  s+1}
{\displaystyle 2}
\right)}{\Gamma\left(\frac{\displaystyle  s}{\displaystyle  2}
\right)}\zeta^{\text{D,N}}_{\text{cyl}}(s+1).
\label{e_4_1}
\end{equation}
It is important to note that here there is no need in additional calculations or
analytic continuation because the  values of
$\zeta_{\text{cir}}^{\text{D,N}}(-1)$,
defining the relevant Casimir energies,
are expressed, according to (\ref{e_4_1}),
in terms of $\zeta_{\text{cyl}}^{\text{D,N}}(0)$.

Let us first derive  $\zeta^{\text{D}}_{\text{cir}}(-1)$ substituting
Eqs.~(\ref{e_3_4}) and (\ref{e_3_9}) into (\ref{e_4_1})
\begin{equation}
\zeta_{\text{cir}}^{\text{D}}(-1)=-\frac{1}{\pi}\lim\limits_{s\to0}^{}
\sum\limits_{i=0}^3 Z_i(s).
\label{e_4_2}
\end{equation}
Numerical calculation and integration according to (\ref{e_3_14}) in
(\ref{e_3_12}) with $s=0$ gives
\begin{equation}
Z_0(0)=-\int_0^{\infty}dy \left[\ln(2 y I_0(y) K_0(y))-
\frac{t^2}{8}(1-6 t^2+5 t^4)\right]=\pi\left(0.02815-\frac{1}{128}\right).
\label{e_4_3}
\end{equation}
In Eq.\ (\ref{e_3_6}) defining $Z_1(s)$ we put $s=0$ and
integrate by parts
\begin{equation}
Z_1(0)=-2\,\zeta(-1)\int_0^{\infty}dy \ln\frac{y}{\sqrt{1+y^2}}
=-2\left(-\frac{1}{12}\right)\left(-\frac{\pi}{2}\right)=-\frac{\pi}{12}.
\label{e_4_4}
\end{equation}
Without pretending to high accuracy the value of $Z_3(0)$ can be evaluated
by making use of Eq.~(\ref{e_3_16})
\begin{eqnarray}
Z_3(0)&=&\frac{1}{32}\,\zeta(3)\,\Gamma\left(\frac{3}{2}\right)
\left[-13\,\Gamma\left(\frac{3}{2}\right)+142\, \Gamma\left(\frac{5}{2}\right)
-\frac{1062}{6}\,\Gamma\left(\frac{7}{2}\right)
+\frac{1356}{24}\,\Gamma\left(\frac{9}{2}\right)\right.\nonumber\\
&&\left. -\frac{565}{720}\,\Gamma\left(\frac{11}{2}\right)
\right] = \frac{\pi}{64}\,(-0.136719)\,\zeta(3)\,{.}
\label{e_4_5}
\end{eqnarray}
The  function $Z_2(s)$ determined in  (\ref{e_3_15}) has a pole
at the point $s=0$ because
\begin{equation}
\zeta(1+s)\simeq\frac{1}{s}+\gamma+\dots,\quad s\to 0\,{.}
\label{e_4_6}
\end{equation}
Therefore, we can only extract the finite and divergent parts in $Z_2(0)$
\begin{equation}
Z_2(0)=\frac{1}{4}\,\Gamma\left(\frac{3}{2}\right)
\Gamma\left(\frac{1}{2}\right) \frac{1}{8}\left(\lim_{s\to0}^{}
\zeta(1+s)+\frac{1}{2}\right)=\frac{\pi}{64}\left(
\left.\frac{1}{s}\,\right|_{s\to0}+\gamma\right)+\frac{\pi}{128}\,{.}
\label{e_4_7}
\end{equation}
Finally, substituting  Eqs.~(\ref{e_4_3}), (\ref{e_4_4}), (\ref{e_4_5}),
and (\ref{e_4_7}) into Eq.~(\ref{e_4_2}) we get
\begin{eqnarray}
\zeta_{\text{cir}}^{\text{D}}(-1)&=&-\frac{1}{\pi}\left[\,0.028156\, \pi +\zeta(-1)\, \pi-
\frac{\pi}{64}\,0.136719 \,\zeta(3)+\frac{\pi}{64}\,
\left(\left.\frac{1}{s}\,\right|_{s\to0}+\gamma\right)\right]
\nonumber\\
&=&\frac{1}{a}\left(0.047189-
\left.\frac{1}{64}\,\frac{1}{s}\,\right|_{s\to0}\right).
\label{e_4_8}
\end{eqnarray}
It gives  the following  result for the Casimir energy  of the scalar
massless field obeying the Dirichlet boundary conditions on a circle
\begin{equation}
E_{\text{cir}}^{\text{D}}=\frac{1}{2}\,\zeta_{\text{cir}}^{\text{D}}(-1)=\frac{1}{a}
\left(0.0023595-\left.
\frac{1}{128}\,\frac{1}{s}\,\right|_{s\to 0} \right).
\label{e_4_9}
\end{equation}

As in the case of an infinite  cylinder, it is convenient first
to construct the sum
of two  zeta functions for the Dirichlet and Neumann boundary conditions
$\zeta_{\text{cir}}^{\text{D+N}}(s)$ and then find
$\zeta_{\text{cir}}^{\text{N}}(s)$
as a difference  $\zeta_{\text{cir}}^{\text{D+N}}(s)
-\zeta_{\text{cir}}^{\text{D}}(s)$.
The zeta function  $\zeta_{\text{cir}}^{\text{D+N}}(s)$ is again expressed
through the corresponding  zeta function of a cylinder
\begin{equation}
\zeta_{\text{cir}}^{\text{D+N}}(s)=2\sqrt{\pi}\,
\frac{\Gamma\left(\frac{\displaystyle s+1}{\displaystyle  2}
\right)}{\Gamma\left(\frac{\displaystyle  s}{\displaystyle  2}\right)}
\zeta_{zyl}^{\text{shell}}(s+1)\,{,}
\label{e_4_10}
\end{equation}
where $\zeta_{\text{cyl}}^{\text{shell}}(s)$ is the zeta function of electromagnetic
field with boundary conditions defined  on a surface of a perfectly
conducting cylindrical shell, $\zeta_{\text{cyl}}^{\text{shell}}(s)=
\zeta_{\text{cyl}}^{\text{D}}(s)+\zeta_{\text{cyl}}^{\text{N}}(s)$.
 The spectral zeta function
$\zeta_{\text{cyl}}^{\text{shell}}(s)$
has been explicitly constructed
in Ref.~\onlinecite{LNB}. The relevant formulas read
\begin{eqnarray}
\zeta_{\text{cyl}}^{\text{shell}}(s)&=&\tilde{Z}_1(s)+ \tilde{Z}_2(s)+\tilde{Z}_3(s),
\label{e_4_11}\\
\tilde{Z}_1(s)&=&\frac{(s-1)\, a^{s-1}}{2\sqrt{\pi}\,
\Gamma\left(\frac{\displaystyle  s}{\displaystyle  2}
\right)\Gamma\left(\frac{\displaystyle  3-s}{\displaystyle  2}\right)}
\int_0^{\infty}dy\, y^{-s} \left\{\ln[1-\mu_0^2(y)]
+\frac{y^4 t^6(y)}{4} \right\},\label{e_4_12}\\
\tilde{Z}_2(s)&=&\frac{(1-s)(3-s)\,a^{s-1}}{64\,\sqrt{\pi}}[2\zeta(s+1)+1]
\,\frac{\Gamma\left(\frac{\displaystyle  1+s}{\displaystyle  2}
\right)}{\Gamma\left(\frac{\displaystyle  s}{\displaystyle  2}\right)},
\label{e_4_13}\\
\tilde{Z}_3(s)&=&\frac{(1-s)(3-s)(71 s^2-52 s-235)\,
a^{s-1}}{61440\sqrt{\pi}}\,\zeta(s+3)\,
\frac{\Gamma\left(\frac{\displaystyle  3+s}{\displaystyle  2}
\right)}{\Gamma\left(\frac{\displaystyle  s}{\displaystyle  2}\right)},
\label{e_4_14}
\end{eqnarray}
where $\mu_0(y) =y(I_0(y)K_0(y))'$.

We are again interested in the value of
$\zeta_{\text{cir}}^{\text{D+N}}(s)$ at the point
$s=-1$ which is expressed through $\zeta_{\text{cyl}}^{\text{shell}}(0)$
according to (\ref{e_4_10}).  The functions $\tilde{Z}_1(s)$
and $\tilde{Z}_3(s)$ give  the finite contributions
to $\zeta_{\text{cir}}^{\text{D+N}}(-1)$
\begin{equation}
-\frac{1}{a}\,0.531627 \quad \mbox{and}\quad \frac{1}{a}\,0.006896\,{,}
\label{e_4_15}
\end{equation}
respectively. However, as in the case of the Dirichlet boundary conditions,  $Z_2(s)$
gives a pole-like contribution to~(\ref{e_4_10})
\begin{equation}
-\frac{3}{32\,a}\left(\left. \frac{1}{s}\,\right|_{s\to0}+\gamma+
\frac{1}{2}\right).
\label{e_4_16}
\end{equation}
Hence, for the Casimir energy in question one obtains
\begin{equation}
E_{\text{cir}}^{\text{D+N}}=\frac{1}{2}\zeta^{\text{D+N}}(-1)=-\frac{3}{64a}
\left.\frac{1}{s}\,\right|_{s \to 0}-\frac{1}{a}\,0.2895.
\label{e_4_17}
\end{equation}
Finding the difference between (\ref{e_4_17}) and (\ref{e_4_9})
we arrive at the Casimir energy of scalar massless field obeying the Neumann
boundary conditions on a circle
\begin{equation}
E_{\text{cir}}^{\text{N}}=\frac{1}{a}\left(-0.3131-
\left.\frac{5}{128}\,\frac{1}{s}\right|_{s\to 0} \right).
\label{e_4_18}
\end{equation}

The zeta function technique does not lead to a finite answer  for the Casimir
energy in the plane problem considered here  (two space-like dimensions),
as well as in all the cases of arbitrary even space dimensions.~\cite{MNg}
As usual the coefficients in front of the pole-like  contributions,
calculated by different methods coincide, but the finite  parts of
the answers   differ.~\cite{L+Romeo,Sen} In this respect our
consideration  has a certain advantage, because when calculating the
Casimir energy of
the fields on a plane we employ Eq.\ (\ref{e_4_1}). Thereby we in fact make
use of
the  spectral zeta function $\zeta_{\text{cyl}}(s)$ for a cylinder which has already
been "normalized"  by a finite answer for the Casimir energy of an infinite
cylinder.

Of course the problem of the Casimir energy calculation in the even
dimensional spaces is far from being completely solved. To obtain
an acceptable finite answer for this energy one should invoke some
additional physical arguments  providing the removal of the pole-like
contributions from Eqs.~(\ref{e_4_9}),  (\ref{e_4_17}), and (\ref{e_4_18})
or  use new mathematical methods which will not result in such
divergent terms.

\section{Conclusion}

In the present paper the explicit expressions  are derived which
define the spectral
zeta functions for an infinite  cylinder $\zeta_{\text{cyl}}(s)$  in a finite
range of complex variable $s$ containing the segment of real axis
$-1\le$ Re~$\,s\le0$. It enables one to find
the spectral
zeta function for a circle making use of the zeta function
for an infinite cylinder according to the relation  (\ref{e_2_6}).
In Ref.~\onlinecite{GR1} this relation was applied directly, i.e.,  for  constructing
$\zeta_{\text{cyl}}(s)$ from $\zeta_{\text{cir}}(s)$. However, to obtain  the value
of $\zeta_{\text{cyl}}(s)$ at the point $s=-1$, the authors  of this paper
have to   make  additional analytic continuation of the
function $\zeta_{\text{cir}}(s)$ from the neighborhood   of the point $s=-1$
to the point $s=-2$.

In our consideration, as well as in treatment of the analogous problems by
other authors,   the central part was played by the uniform
asymptotic expansion for the product of the Bessel functions.
The lack  of such expansions  for the eigenfunctions in problems
with other geometry of the boundaries (for example, with
boundary conditions   defined on the surface of a spheroid)
does not permit to expand directly this approach beyond
the systems with spherical symmetry.

\acknowledgments

The authors are indebted to Michael Bordag (ITP, Leipzig University)
for many valuable
discussions of the problems considered in this paper.
The work was accomplished  with financial support of  the
Heisenberg--Landau Program (Grant No.~99-11) and Russian Foundation for
Basic Research (Grant No.~97-01-00745).


\begin{thebibliography}{99}
\bibitem{endnote} In the next section it will be shown that
$\zeta_{\text{cyl}}^{\text{D}}(s)$ has a simple pole at the point $s=0$.
\end{thebibliography}
\end{document}